\documentstyle[12pt,epsf]{article} 

\def\lsim{\mathrel{\lower2.5pt\vbox{\lineskip=0pt\baselineskip=0pt 
           \hbox{$<$}\hbox{$\sim$}}}} 
\def\gsim{\mathrel{\lower2.5pt\vbox{\lineskip=0pt\baselineskip=0pt 
           \hbox{$>$}\hbox{$\sim$}}}}

\begin{document} 

\begin{flushright}
DPNU-00-37 \\ hep-ph/0011004
\end{flushright}

\vspace{10mm}

\begin{center}
{\Large \bf
The $\mu$-Problem and Seesaw-type Mechanism\\ in the Higgs Sector}

\vspace{20mm}

Masato ITO 
\footnote{E-mail address: mito@eken.phys.nagoya-u.ac.jp}
\end{center}

\begin{center}
{
\it 
{}Department of Physics, Nagoya University, Nagoya, 
JAPAN 464-8602 
}
\end{center}

\vspace{25mm}

\begin{abstract}
We explore a new solution to the $\mu$-problem.
In the scenario of SUSY-breaking mediation through anti-generation fields, 
we find that the $B\mu$ term has its origin in a
seesaw-type mechanism as well as in a loop diagram through gauge
interactions.
It is shown that the dominant contributions to the $B\mu$ term 
are controlled by the flavor symmetry in the model.
\end{abstract}

\newpage 
\section{Introduction}

Supersymmetry (SUSY) is expected to become an important ingredient of new
physics beyond the Standard Model. 
In particular, SUSY provides the most attractive
mechanism to stabilize the scale hierarchy between the Planck 
scale, $M_{Pl}$, 
and the electroweak scale $M_{w}$.
Because we have not yet observed superparticles up to the order of 
$M_{w}$, SUSY should be broken at scales above $M_{w}$, but not far 
from $M_{w}$.
The phenomenology of SUSY depends on the mechanism of SUSY-breaking and
on the manner in which SUSY-breaking in the hidden sector 
is transmitted to the observable sector.

Recently, in Ref. \cite{ourmodel}, a new scenario of SUSY-breaking
mediation was proposed in the
framework of a weakly-coupled heterotic string on the Calabi-Yau
compactification.
In this scenario, the breakdown of supersymmetry in the
hidden sector is transmitted to anti-generation fields by
gravitational interactions, and subsequent transmission of the breaking
to the observable sector occurs through gauge interactions.
It has been shown that the spectra of the superparticles
given by the model are phenomenologically viable.

In the MSSM, we are confronted with the $\mu$-problem.
The supersymmetric $\mu$ term is the only dimensional
parameter in the superpotential
%
%%%%%%%%%%%%%%%%%%%%% Eqn.1 %%%%%%%%%%%%%%%%%%%%%%%%%%%%%%
 \begin{equation}
  W_{\rm MSSM}\supset\mu H_{u}H_{d}\,.
 \label{eqn1}
 \end{equation}
%%%%%%%%%%%%%%%%%%%%%%%%%%%%%%%%%%%%%%%%%%%%%%%%%%%%%%%%%%
%
The Higgs scalar potential is given by
%
%%%%%%%%%%%%%%%%%%%%%% Eqn.2 %%%%%%%%%%%%%%%%%%%%%%%%%%%%%%
 \begin{eqnarray}
  V_{\rm Higgs}&=&m^{2}_{1}|H_{u}|^{2}+m^{2}_{2}|H_{d}|^{2}
                 -\left(B\mu H_{u}H_{d}+{\rm h.c}\right)\nonumber\\
  && +\frac{g^{2}_{2}}{8}\left(H^{\dagger}_{d}\vec{\sigma}H_{d}
                              +H^{\dagger}_{u}\vec{\sigma}H_{u}\right)^{2}
     +\frac{g^{2}_{1}}{8}
      \left(|H_{u}|^{2}-|H_{d}|^{2}\right)^{2}
  \label{eqn2}
 \end{eqnarray}
%%%%%%%%%%%%%%%%%%%%%%%%%%%%%%%%%%%%%%%%%%%%%%%%%%%%%%%%%%%
%
with $m^{2}_{1}=m^{2}_{H_{u}}+\mu^{2}$ and
$m^{2}_{2}=m^{2}_{H_{d}}+\mu^{2}$, 
where $m_{H_{u}}$ and $m_{H_{d}}$ are soft SUSY-breaking
Higgs masses, and
$g_{1}$ and $g_{2}$ are $U(1)$ and $SU(2)$ gauge coupling constants,
respectively.
The $B\mu$ term is the soft SUSY-breaking term for Higgs bilinear couplings.
The two Higgs fields acquire the vacuum expectation values (VEVs)
%%%%%%%%%%%%%%%%%%%%%%%%%%% Eqn.3 %%%%%%%%%%%%%%%%%%%%%%%%%%%%%%%%
\begin{equation}
 \langle H_{u}\rangle
 =\left(\begin{array}{c}0\\v_{u}\end{array}\right)\,,\hspace{1cm}
 \langle H_{d}\rangle
 =\left(\begin{array}{c}v_{d}\\0\end{array}\right)\label{eqn3}\,,
\end{equation}
%%%%%%%%%%%%%%%%%%%%%%%%%%%%%%%%%%%%%%%%%%%%%%%%%%%%%%%%%%%%%%
with $v^{2}_{u}+v^{2}_{d}\simeq (174\;{\rm GeV})^{2}$.
The minimization condition of the scalar potential leads to the relations
%%%%%%%%%%%%%%%%%%%%%%%%%% Eqn.4,5 %%%%%%%%%%%%%%%%%%%%%%%%%%%%%%%%%
\begin{eqnarray}
 \frac{M^{2}_{Z}}{2}&=&-\mu^{2}+
                   \frac{m^{2}_{H_{d}}-m^{2}_{H_{u}}\tan^{2}\beta}
                        {\tan^{2}\beta -1}\,,\label{eqn4}\\
 2B\mu&=&\left(2\mu^{2}+m^{2}_{H_{d}}+m^{2}_{H_{u}}\right)
           \sin 2\beta\,,\label{eqn5}
\end{eqnarray}
%%%%%%%%%%%%%%%%%%%%%%%%%%%%%%%%%%%%%%%%%%%%%%%%%%%%%%%%%%%%%%%%%%%
with $\tan\beta=v_{u}/v_{d}$.
The electroweak symmetry breaking occurs naturally in the MSSM, because
$m^{2}_{H_{u}}$ becomes negative due to the dominant contribution
of top Yukawa coupling $h_{t}$ in the RG evolution as
%%%%%%%%%%%%%%%%%%%%%% Eqn.6 %%%%%%%%%%%%%%%%%%%%%%%%%%%%%%%%%%%%%%%
\begin{equation}
 m^{2}_{H_{u}}(M_{Z})=m^{2}_{H_{d}}(M_{Z})
                      -\frac{6h^{2}_{t}}{16\pi^{2}}m^{2}_{\tilde{t}}
                      \log\left(\frac{\Lambda_{RG}}{M_{Z}}\right)^{2}\,.
 \label{eqn6}
\end{equation}
%%%%%%%%%%%%%%%%%%%%%%%%%%%%%%%%%%%%%%%%%%%%%%%%%%%%%%%%%%%%%%%%%%%%
Here $m_{\tilde{t}}$ is the stop mass and
$\Lambda_{RG}$ is the renormalization scale.
For $\tan\beta\gsim 2$, Eq. (\ref{eqn4}) becomes
%%%%%%%%%%%%%%%%%%%%%% Eqn.7 %%%%%%%%%%%%%%%%%%%%%%%%%%%%%%%%%%%%%%
\begin{equation}
 \frac{M^{2}_{Z}}{2}\sim -\mu^{2}-m^{2}_{H_{u}}\label{eqn7}\,.
\end{equation}
%%%%%%%%%%%%%%%%%%%%%%%%%%%%%%%%%%%%%%%%%%%%%%%%%%%%%%%%%%%%%%%%%%%
Therefore, phenomenologically, the values of $\mu$ and $B\mu$
should satisfy the condition
%
%%%%%%%%%%%%%%%%%%%%%% Eqn.8 %%%%%%%%%%%%%%%%%%%%%%%%%%%%%%
 \begin{equation}
  \mu^{2}\sim B\mu={\cal O}((100\;{\rm GeV})^{2})\,.
   \label{eqn8}
 \end{equation}
%%%%%%%%%%%%%%%%%%%%%%%%%%%%%%%%%%%%%%%%%%%%%%%%%%%%%%%%%%%
%
An important question in the mediation of SUSY-breaking is
whether or not we can obtain $\mu$ and $B\mu$ terms to be consistent
with Eq. (\ref{eqn8}). 
This is the so-called $\mu$-problem.
We would like to determine the dynamical mechanism by which
$\mu$ and $B\mu$ terms are generated.
%%%%%%%%%%%%%%%%%%%%%% Fig.1 %%%%%%%%%%%%%%%%%%%%%%%%%%%%%%
\begin{figure}
      \epsfxsize=6cm
\centerline{\epsfbox{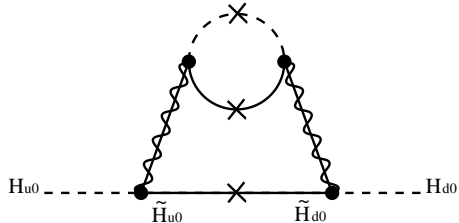}}
\caption{Diagram for generating the $B\mu$ term a via two-loop diagram.}
\label{fig1} 
\end{figure}
%%%%%%%%%%%%%%%%%%%%%%%%%%%%%%%%%%%%%%%%%%%%%%%%%%%%%%%%%%%
In the model studied in Ref. \cite{ourmodel}, the $B\mu$ term is generated by
a two-loop diagram through gauge interactions, as shown in Fig. \ref{fig1},
where anti-generation fields propagate the inner loop, and the remaining
internal line has the fermion mass insertion, the $\mu$ term.
It has been shown that this model exhibits a phenomenologically
viable solution to the $\mu$-problem. 

We now give some comments on the $\mu$-problem.
The $\mu$ and $B\mu$ terms are of different characters.
Since the $\mu$ term is supersymmetric, it cannot be generated by
quantum effects, as shown by the non-renormalization theorem.
Namely, $\mu$ arises only at tree level.
On the other hand, since the $B\mu$ is a soft SUSY-breaking term,
it can be generated by quantum effects associated with SUSY-breaking.

For example, a simple model for generating these terms has been proposed
in the gauge-mediation scenario.
In this model, the superpotential contains the term $XH_{u}H_{d}$,
where $X$ is the singlet superfield.
When $X$ develops a nonzero VEV, i.e., 
$\langle X\rangle=X+\theta^{2}F^{X}$, then we obtain
$\mu\sim X$ and $B\mu\sim F^{X}$.
In this scenario, the gaugino mass is given by 
$m_{\lambda}\sim(g^{2}/16\pi^{2})F^{X}/X$.
When we take $m_{\lambda}\sim\mu={\cal O}(100\;{\rm GeV})$, 
this implies $B\mu\gg \mu^{2}$,
which is phenomenologically unacceptable.
This is because the $\mu$ and $B\mu$ terms have their origin in the
same coupling.
Thus, it is difficult to avoid this problem,
unless we introduce a new symmetry and additional fields by hand. \cite{mu}
However, we choose not to introduce new symmetry and additional fields
only for the purpose of solving the $\mu$-problem. 
Rather, we rely on the flavor symmetry, 
which is favorable for explaining the quark/lepton mass hierarchy.
In this situation, $\mu$ and $B\mu$ are represented by independent
parameters.
As discussed below, we show that the magnitude of each term is controlled by
flavor symmetry.

For supergravity to be a low energy effective theory of superstrings, 
all soft terms must depend strongly on the form of the K\"ahler potential.
Although it is difficult to study the definite form
of the K\"ahler potential in generic Calabi-Yau compactification,
typical forms are known in special cases, such as 
the large radius limit and orbifold compactification 
($Z_{N}$ or $Z_{N}\times Z_{M}$),
and the relation $B\mu\sim\mu^{2}\sim(m_{3/2})^{2}$
can be derived by approximating the scalar potential.
For this reason, approaches to the $\mu$-problem have been tried only 
by carrying out suitable modification of forms of the K\"ahler potential.
In contrast to such approaches,
assuming that K\"ahler potential is of the minimal form,
we explore an alternative approach
to the $\mu$-problem.
%
%Recently, by introducing new symmetry and additional fields, the 
%other solutions to the $\mu$-problem had been proposed \cite{mu}.

\section{Mechanism for generating ${\rm B\mu}$ term}

In this paper we consider the scenario in which SUSY-breaking is 
mediated by anti-generation fields, proposed in Ref. \cite{ourmodel}, and
discuss the possibility of generating a $B\mu$ term with a 
seesaw-type mechanism as well as with a loop diagram through 
gauge interactions, as shown in Fig. \ref{fig1}.
Further, we study the $\mu$-problem in this scenario.

The model discussed here is based on a string-inspired model. \cite{E6a}
The effective theory in the observable sector from the
weakly-coupled Calabi-Yau string is characterized by $N=1$ SUSY, the $E_{6}$
gauge group, and massless matter fields which belong to ${\bf 27}$ and
${\bf 27}^{\ast}$ representations in $E_{6}$.
The massless chiral superfields consist of
%
%%%%%%%%%%%%%%%%%%%%%%%%%%% Eqn.9 %%%%%%%%%%%%%%%%%%%%%%
 \begin{equation}
 N_{f}\;\Phi({\bf 27})+
 \delta\left(\Phi_{0}({\bf 27})+\overline{\Phi}({\bf 27^{\ast}})\right)\,,
  \label{eqn9}
 \end{equation}
%%%%%%%%%%%%%%%%%%%%%%%%%%%%%%%%%%%%%%%%%%%%%%%%%%%%%%%%%%%
%
where $N_{f}$ represents the family number at low energies.
Here, $\delta$ sets of vector-like multiplets are included in the massless
sector.
The numbers $N_{f}+\delta$ and $\delta$ are the generation
number and anti-generation number, respectively.
We set $N_{f}=3$ and $\delta =1$ for the sake of simplicity.
In the effective theory, in general, there appear certain discrete 
symmetries $G_{\rm st}$ as a stringy selection rule associated with
symmetric structure of the compactified space. \cite{E6a,E6b}
Let us consider the case $G_{\rm st}=G_{\rm fl}\times Z_{2}$, 
where $G_{\rm fl}$ is the flavor symmetry and $Z_2$ is the $R$-parity. 
The $R$-parity distinguishes $\Phi_{i}\;(i=1,2,3)$ from a vector-like set. 
It is assumed that ordinary quarks and leptons are included in the chiral
multiplets $\Phi_{i}\;(i=1,2,3)$ and that the $R$-parity of all
$\Phi_{i}$ are odd. 
Because light Higgs scalars are even under $R$-parity, light Higgs doublets
must reside in the set $\{\Phi_{0},\overline{\Phi}\}$, whose
elements are vector-like multiplets.
For this reason we assign even $R$-parity to $\Phi_{0}$ and
$\overline{\Phi}$.
Hence, through the spontaneous breaking of gauge symmetry, gauge
superfields possibly mix with the vector-like multiplets
$\Phi_{0}$ and $\overline{\Phi}$ but not mix with the chiral multiplets
$\Phi_{i}\;(i=1,2,3)$.
No mixing occurs between the vector-like multiplets and the
chiral multiplets.
Hence, in the low energy region, ordinary quarks and leptons arise from
$\Phi_{i}\;(i=1,2,3)$ and ordinary Higgs fields from $\Phi_{0}$. 
Moreover, there appear a dilaton field $D$, K$\ddot{a}$hler class 
moduli fields $T_{i}$, and complex structure moduli fields $U_{i}$.
The VEV of the dilaton field $D$ determines the gauge coupling constant
and the VEVs of the moduli fields $U_{i}$ and $T_{i}$ parametrize the size
and shapes of the compactified manifold.

We choose $SU(6)\times SU(2)_{R}$ as a typical example of
the unification gauge group. 
With this gauge group, chiral superfields ($\Phi$) in the ${\bf 27}$
representation of $E_{6}$ are decomposed as
%
%%%%%%%%%%%%%%%%%%%%%%%%% Eqn.10 %%%%%%%%%%%%%%%%%%%%%%%%%
 \begin{eqnarray}
   E_{6}&\supset&  SU(6)  \times  SU(2)_{R} \nonumber\\
    &&(\hspace{3mm}{\bf 15}\hspace{4mm} , \hspace{4mm}{\bf 1}\hspace{3mm})  
    \hspace{3mm}Q,L,g,g^{c},S\,,\nonumber\\
    &&(\hspace{4mm}{\bf 6^{\ast}}\hspace{3.5mm} ,\hspace{4mm} {\bf 2}
    \hspace{3mm})\hspace{3mm}
  (U^{c},D^{c}),(N^{c},E^{c}),(H_{u},H_{d}).
%  \end{tabular}
   \label{eqn10}
 \end{eqnarray}
%%%%%%%%%%%%%%%%%%%%%%%%%%%%%%%%%%%%%%%%%%%%%%%%%%%%%%%%%%%
%
The $SU(6)\times SU(2)_{R}$ model has the following attractive features.
\cite{E6b}
The first feature is that without introducing additional adjoint fields,
the gauge symmetry is spontaneously broken to the Standard Model gauge group
in two steps at scales 
$\langle S_{0}\rangle=\langle \overline{S}\rangle\sim 10^{17}\;{\rm GeV}$ and 
$\langle N_{0}\rangle=\langle \overline{N^{c}}\rangle\sim 10^{16}\;{\rm GeV}$. 
The second feature is that
this model is free from triplet-doublet splitting problem, because the
light Higgs doublets $H_{u}$ and $H_{d}$ and the color-triplet Higgs
$g$ and $g^{c}$ belong to different irreducible representations.

It is assumed that the SUSY-breaking resulting from gaugino
condensation in the hidden sector is only transmitted to
the $F$-component of the moduli field $T$.
Specifically, the moduli field $T$ develops a SUSY-breaking VEV, 
$(\langle T\rangle =T+\theta^{2}F^{T})$, while
both $D$ and $U$ have vanishing $F$-components $(F^{D}=F^{U}=0)$.
This situation is the same as that in the moduli-dominated SUSY-breaking
scenario. \cite{moduli}
At present, it is unknown how to specify the seeds of SUSY-breaking 
through non-perturbative dynamics in superstring theory.
Therefore, we make an assumption regarding the source of SUSY-breaking.

The superpotential $W$ in the observable sector
is described in terms of $\Phi({\bf 27})$ and
$\overline{\Phi}({\bf 27^{\ast}})$. 
Since we are interested in the $\mu$-problem, we concentrate our 
attention on the terms incorporating $SU(2)_{L}$-doublet Higgs, 
%%%%%%%%%%%%%%%%%%%%% Eqn.11 %%%%%%%%%%%%%%%%%%%%%%%%%%%%%%%%%
 \begin{eqnarray}
  W & \sim & 
    h(T,\,U) \left( \frac{S_0 \overline{S}}{M_s^2} \right)^s S_0H_{u0}H_{d0} 
  + f(T,\,U) \left( \frac{S_0 \overline{S}}{M_s^2} \right)^{\overline s} 
            \overline{S}\;\overline{H_{u}}\;\overline{H_{d}} \nonumber \\
   & & + g(T,\,U) M_s \left( \frac{S_{0}\overline{S}}{M_s^2} \right)^b 
         \left( H_{u0}\overline{H_{u}} + 
              H_{d0}\overline{H_{d}} \right)\,,  
 \label{eqn11}
 \end{eqnarray}
%%%%%%%%%%%%%%%%%%%%%%%%%%%%%%%%%%%%%%%%%%%%%%%%%%%%%%%%%%%%%%
where the exponents $s$, $\overline{s}$ and $b$ are non-negative 
integers, which are governed by the flavor symmetry. 
Hereafter we assume $h(T, \,U), \ f(T, \,U), \ g(T, \,U)\sim{\cal O}(1)$.
For a Calabi-Yau string, generation fields and anti-generation 
fields in the trilinear terms couple separately to the $U$ moduli and 
the $T$ moduli, respectively.
In view of this feature of the trilinear terms,
we assume that $h(T, \,U)$ and $f(T, \,U)$ depend only on the complex 
structure moduli $U$ and the K\"ahler moduli $T$, respectively. 

When $S_{0}$ and $\overline{S}$ develop nonzero VEVs, 
the above superpotential induces the bilinear terms 
%%%%%%%%%%%%%%%%%%%%%%%%%% Eqn.12 %%%%%%%%%%%%%%%%%%%%%%%%%%%%%
\begin{equation}
 W_{h}= h(U) \, x^s \langle S_{0}\rangle H_{u0}H_{d0} + 
        f(T) \, x^{\overline s} 
          \langle\overline{S}\rangle\overline{H_{u}}\;\overline{H_{d}}
     + m\left( \;H_{u0}\overline{H_{u}}+H_{d0}\overline{H_{d}}\; \right)\,,
\label{eqn12}
\end{equation}
%%%%%%%%%%%%%%%%%%%%%%%%%%%%%%%%%%%%%%%%%%%%%%%%%%%%%%%%%%%%%%%
where 
%%%%%%%%%%%%%%%%%%%%%%% Eqn.13 %%%%%%%%%%%%%%%%%%%%%%%%%%%%%%%%%
 \begin{equation}
 x =\frac{\langle S_{0}\rangle \langle\overline{S}\rangle }{M_s^2}
   ={\cal O}(10^{-2})
 \label{eqn13}
 \end{equation}
%%%%%%%%%%%%%%%%%%%%%%%%%%%%%%%%%%%%%%%%%%%%%%%%%%%%%%%%%%%%%%%%%
and 
%%%%%%%%%%%%%%%%%%%%%%%%%%%% Eqn.14 %%%%%%%%%%%%%%%%%%%%%%%%%%%%
\begin{equation}
  m = g(T, \,U) \, M_s \, x^b. 
\label{eqn14}
\end{equation}
%%%%%%%%%%%%%%%%%%%%%%%%%%%%%%%%%%%%%%%%%%%%%%%%%%%%%%%%%%%%%%%%%
We now consider the $\mu$ term, which is the first term on the 
r.h.s. of Eq. (\ref{eqn12}). 
From this we obtain
%%%%%%%%%%%%%%%%%%%%%%%%%% Eqn.15 %%%%%%%%%%%%%%%%%%%%%%%%%%%%%
\begin{equation}
   \mu =h(U) \, x^s \langle S_{0}\rangle\,.
\label{eqn15}
\end{equation}
%%%%%%%%%%%%%%%%%%%%%%%%%%%%%%%%%%%%%%%%%%%%%%%%%%%%%%%%%%%%%%%%%%%%%%
For an appropriate value of $s$, the $\mu$ term can be taken to be on the
electroweak scale.\cite{E6b}
The choice of the exponent $s$ is linked to the flavor charge assignment for 
$\Phi_{0}({\bf 15,1})$, $\Phi_{0}({\bf 6^{\ast},2})$, 
$\overline{\Phi}({\bf 15,1})$ and $\overline{\Phi}({\bf 6^{\ast},2})$. 
For the second term in Eq. (\ref{eqn11}), we assume 
$\overline{s} = 0$. 
The function $f(T)$ is determined by the string worldsheet 
instanton effects \cite{instanton} as 
%%%%%%%%%%%%%%%%%%%%%%%%%%%%% Eqn.16 %%%%%%%%%%%%%%%%%%%%%%%
\begin{equation}
f(T)\propto e^{-2\pi T/M_{s}}\,.
\label{eqn16}
\end{equation}
%%%%%%%%%%%%%%%%%%%%%%%%%%%%%%%%%%%%%%%%%%%%%%%%%%%%%%%%%%%%

Next, we turn to the seesaw-type mechanism of generating the $B\mu$ term. 
From Eq. (\ref{eqn12}), we have the $4\times 4$ mass-squared matrix
%%%%%%%%%%%%%%%%%%%%%%%%% Eqn.17 %%%%%%%%%%%%%%%%%%%%%%%%%%%%%%%%%%%%%
 \begin{equation}
  M^{2}=
  \begin{array}{cl}
    & \hspace{1.2cm}\overline{H_{d}}\hspace{1.6cm}
      \overline{H_{u}}^{\dagger}\hspace{1.8cm}
      H_{d0}^{\dagger}\hspace{1.6cm}
      H_{u0}\\
    \begin{array}{c}
    \overline{H_{d}}^{\dagger}\\ \overline{H_{u}}\\
  H_{d0}\\ H_{u0}^{\dagger}
    \end{array}
    & \left(\begin{array}{cccc}
      |p|^{2}+|m|^{2} & q^{\ast} & 0 & p^{\ast}m+\mu m^{\ast}\\
     q & |p|^{2}+|m|^{2} & pm^{\ast}+\mu^{\ast}m & 0 \\ 
     0 & p^{\ast}m+\mu m^{\ast} & |m|^{2}+|\mu|^{2} & 0\\
     pm^{\ast}+\mu^{\ast}m & 0 & 0 & |m|^{2}+|\mu|^{2}
      \end{array}\right)
  \end{array}
  \label{eqn17}
 \end{equation}
%%%%%%%%%%%%%%%%%%%%%%%%%%%%%%%%%%%%%%%%%%%%%%%%%%%%%%%%%%%%%%%%%%%%%
for Higgs scalar fields, where $p$ and $q$ are defined as
%%%%%%%%%%%%%%%%%%%%%%%%%%% Eqn.18,19 %%%%%%%%%%%%%%%%%%%%%%%%%%
\begin{eqnarray}
 p&=&f(T)\langle\overline{S}\rangle\label{eqn18}\,,\\
 q&=&f^{\prime}(T)F^{T}\langle\overline{S}\rangle\label{eqn19}\,.
\end{eqnarray}
%%%%%%%%%%%%%%%%%%%%%%%%%%%%%%%%%%%%%%%%%%%%%%%%%%%%%%%%%%%%%%%%%%%%%
In Eq. (\ref{eqn17}), $p$, $q$, $\mu$ and $m$ can be complex numbers.
Here $f^{\prime}(T)$ represents the derivative of $f(T)$ with respect to 
the moduli $T$.
The magnitude of $p$ becomes 
%%%%%%%%%%%%%%%%%%%%%%%%%%% Eqn.20 %%%%%%%%%%%%%%%%%%%%%%%%
\begin{equation}
|p|=|f(T)\langle S_{0}\rangle|\sim{\cal O}(10^{17}\;{\rm GeV})
\label{eqn20}
\end{equation}
%%%%%%%%%%%%%%%%%%%%%%%%%%%%%%%%%%%%%%%%%%%%%%%%%%%%%%%%%%%%
%%%%%%%%%%%%%%%%%%%%%%%% Fig.2 %%%%%%%%%%%%%%%%%%%%%%%%%%%%%%%
\begin{figure}
      \epsfxsize=8cm
\centerline{\epsfbox{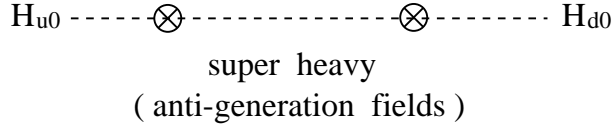}}
\caption{Diagram for generation of the $B\mu$ term via 
the seesaw-type mechanism.}
\label{fig2} 
\end{figure}
%%%%%%%%%%%%%%%%%%%%%%%%%%%%%%%%%%%%%%%%%%%%%%%%%%%%%%%%%%%%%%
The scale $m$ depends on the exponent $b$
and becomes smaller than $\langle S_{0}\rangle$.
Here we assume
%%%%%%%%%%%%%%%%%%%%%%%%%%%%% Eqn.21 %%%%%%%%%%%%%%%%%%%%%%%%%%%%%
 \begin{equation}
  |p|\gg|m|\gg|\mu|\sim{\cal O}(100\;{\rm GeV})\,.\label{eqn21}
 \end{equation}
%%%%%%%%%%%%%%%%%%%%%%%%%%%%%%%%%%%%%%%%%%%%%%%%%%%%%%%%%%%%%%%
%%%%%%%%%%%%%%%%%%%%%%%% Fig.2 %%%%%%%%%%%%%%%%%%%%%%%%%%%%%%%
%\begin{figure}
%      \epsfxsize=8cm
%\centerline{\epsfbox{seesawBmu.eps}}
%\caption{Diagram for generation of the $B\mu$ term via 
%the seesaw-type mechanism.}
%\label{fig2} 
%\end{figure}
%%%%%%%%%%%%%%%%%%%%%%%%%%%%%%%%%%%%%%%%%%%%%%%%%%%%%%%%%%%%%%
Since the anti-generation fields become superheavy,
they decouple at low energies.
Further, we assume $|F^{T}|\ll M^{2}_{s}$.
Then we have $|p|\gg \sqrt{|q|}$.

Although we have no Higgs bilinear coupling, i.e., the $H_{u0}$-$H_{d0}$ and
$H^{\dagger}_{u0}$-$H^{\dagger}_{d0}$ components of $M^{2}$ vanish,
after integrating out the heavy modes $\overline{H_{u}}$ and 
$\overline{H_{d}}$, Higgs bilinear 
couplings are induced at low energies due to the seesaw-type mechanism.
Concretely, a $B\mu$ term is induced through
propagation of the heavy anti-generation fields 
$\overline{H_{u}}$ and $\overline{H_{d}}$, as shown in Fig. \ref{fig2},
in which the dotted line between the circled crosses represents 
a linear combination of the heavy anti-generation scalar fields 
$\overline{H_{u}}$ and $\overline{H_{d}}$.
The matrix Eq. (\ref{eqn17}) is equivalently represented by the 
block matrix 
%%%%%%%%%%%%%%%%%%%%%%%%% Eqn.22 %%%%%%%%%%%%%%%%%%%%%%%%%%%%%%%%%%%
 \begin{equation}
  M^{2}=
  \left(
        \begin{array}{cc}
         X & Y \\ & \\ Y & Z
        \end{array}
  \right)\,,
  \label{eqn22}
 \end{equation}
%%%%%%%%%%%%%%%%%%%%%%%%%%%%%%%%%%%%%%%%%%%%%%%%%%%%%%%%%%%%%%%%%
with
%%%%%%%%%%%%%%%%%%%%%%%%% Eqn.23,24,25 %%%%%%%%%%%%%%%%%%%%%%%%%%
\begin{eqnarray}
X &=& \left(\begin{array}{cc}
            |p|^{2}+|m|^{2} & q^{\ast}\\
            q & |p|^{2}+|m|^{2}
            \end{array}\right)\,,\label{eqn23}\\
Y &=& \left(\begin{array}{cc}
            0 & p^{\ast}m+\mu m^{\ast}\\
            pm^{\ast}+\mu^{\ast}m & 0
            \end{array}\right)\,,\label{eqn24}\\
Z &=& \left(|m|^{2}+|\mu|^{2}\right)
      \left(\begin{array}{cc}
            1 & 0 \\ 0 & 1
            \end{array}\right)\label{eqn25}\,.
\end{eqnarray}
%%%%%%%%%%%%%%%%%%%%%%%%%%%%%%%%%%%%%%%%%%%%%%%%%%%%%%%%%%%%%%%%%

In comparison with the usual gauge mediation scenario,\cite{GMSB} 
the seesaw-type mechanism depicted in Fig. \ref{fig2} is worthy of note.
The anti-generation fields $\overline{H_{u}}$ and $\overline{H_{d}}$
behave as the messenger fields in the usual gauge mediation,
in which scenario messenger fields do not directly couple to the
ordinary Higgs fields.
Therefore, in the usual gauge mediation scenario we have vanishing $Y$
for the mass matrix Eq. (\ref{eqn22}) and the $B\mu$ term arises only
from the two-loop diagram illustrated in Fig. \ref{fig1}.
By contract, in the present model, the sub-matrix $Y$ is non-vanishing
and causes the seesaw-type mechanism to occur. 
Because the anti-generation fields couple to the observed Higgs 
fields through the non-renormalizable interactions,
we need to take account of the seesaw-type mechanism 
as well as the two-loop diagram through gauge interactions.

The mass-squared values of the heavy anti-generation Higgs fields are
given approximately by the sub-matrix $X$, 
and the eigenvalues become $|p|^{2}+|m|^{2}\pm |q|$.
The mass splitting in the heavy modes can be attributed to the
SUSY-breaking source $F^{T}$.

For light modes, the mass-squared values are obtained after
integrating out the heavy modes.
Explicitly, we have
%%%%%%%%%%%%%%%%%%%%%%%%%%%%% Eqn.26 %%%%%%%%%%%%%%%%%%%%%%%%%%%
\begin{equation}
    Z-YX^{-1}Y=
     \left(\begin{array}{ccc}
      x & &y \\
        & &\\
      y^{\ast} & &x
     \end{array}\right)\,,
  \label{eqn26}
\end{equation}
%%%%%%%%%%%%%%%%%%%%%%%%%%%%%%%%%%%%%%%%%%%%%%%%%%%%%%%%%%%%%%
where $x$ and $y$ are written as
%%%%%%%%%%%%%%%%%%%%%%%%%%%% Eqn.27,28 %%%%%%%%%%%%%%%%%%%%%%%
\begin{eqnarray}
 x &=& |m|^{2}+|\mu|^{2}-\frac{|p|^{2}+|m|^{2}}
                            {(|p|^{2}+|m|^{2})^{2}-|q|^{2}}
                            |p^{\ast}m+\mu m^{\ast}|^{2}\label{eqn27}\,,\\
 y &=& \frac{q(p^{\ast}m+\mu m^{\ast})^{2}}
            {(|p|^{2}+|\mu|^{2})^{2}-|q|^{2}}\label{eqn28}\,.
\end{eqnarray}
%%%%%%%%%%%%%%%%%%%%%%%%%%%%%%%%%%%%%%%%%%%%%%%%%%%%%%%%%%%%%%%%%
The off-diagonal elements $y$ and $y^{\ast}$ result from the SUSY-breaking 
$F^{T}$ and represent $B\mu$ and $B\mu^{\ast}$.
From Eq. (\ref{eqn21}), the quantity $B\mu^{(\rm seesaw)}$ generated
by the seesaw-type mechanism can be approximated as
%%%%%%%%%%%%%%%%%%%%%%%%%%%%% Eqn.29 %%%%%%%%%%%%%%%%%%%%%%%%%%%
 \begin{equation}
  B\mu^{(\rm seesaw)}=y\sim q\frac{m^{2}}{p^{2}}\,.
  \label{eqn29}
 \end{equation}
%%%%%%%%%%%%%%%%%%%%%%%%%%%%%%%%%%%%%%%%%%%%%%%%%%%%%%%%%%%%%%%%%
The ratio of $B\mu^{(\rm seesaw)}$ to $\mu^{2}$ becomes
%%%%%%%%%%%%%%%%%%%%%%%%%%%%%% Eqn.30 %%%%%%%%%%%%%%%%%%%%%%%%%%%%%%
 \begin{equation}
  \frac{B\mu^{(\rm seesaw)}}{\mu^{2}}=
  \frac{1}{{\cal O}(10^{3}\;{\rm GeV}^{2})}
  \times \frac{\Lambda m^{2}}{p}\,,
  \label{eqn30}
 \end{equation}
%%%%%%%%%%%%%%%%%%%%%%%%%%%%%%%%%%%%%%%%%%%%%%%%%%%%%%%%%%%%%%%%%%%%%
where we take $\mu\sim{\cal O}(100\;{\rm GeV})$.
Here we have used $q=2\pi p\Lambda$, and $\Lambda$ is defined as
%%%%%%%%%%%%%%%%%%%%%%%%%%%%%%% Eqn.31 %%%%%%%%%%%%%%%%%%%%%%%%%%%%%%
 \begin{equation}
  \Lambda=\frac{F^{T}}{M_{s}}\,.\label{eqn31}
 \end{equation}
%%%%%%%%%%%%%%%%%%%%%%%%%%%%%%%%%%%%%%%%%%%%%%%%%%%%%%%%%%%%%%%%%%%%%

As mentioned above, in the present model, the $B\mu$ term is also 
generated by the two-loop diagram through gauge interactions, as shown 
in Fig. \ref{fig1}.
The contribution of loop diagram is given by \cite{ourmodel}
%%%%%%%%%%%%%%%%%%%%%%%%%%%%%%% Eqn.32 %%%%%%%%%%%%%%%%%%%%%%%%%%%%%%
 \begin{equation}
  \frac{B\mu^{(\rm loop)}}{\mu^{2}}= {\cal O}(10)\times
  \frac{g^{2}_{2}}{16\pi^{2}}\frac{m_{\lambda_{2}}}{\mu}
  \ln\left(\frac{m_{\lambda_{2}}}{\mu}\right)
  =\frac{1}{{\cal O}(10^{4}\;{\rm GeV})}\times \Lambda\,,\label{eqn32}
 \end{equation}
%%%%%%%%%%%%%%%%%%%%%%%%%%%%%%%%%%%%%%%%%%%%%%%%%%%%%%%%%%%%%%%%%%%%%
where the $SU(2)$ gaugino mass $m_{\lambda_{2}}$ comes from the
one-loop diagram of the anti-generation fields and is expressed as 
\cite{ourmodel}
%%%%%%%%%%%%%%%%%%%%%%%%%%%%%%% Eqn.33 %%%%%%%%%%%%%%%%%%%%%%%%%%%%%%
 \begin{equation}
 m_{\lambda_{2}}\sim\frac{g^{2}_{2}}{8\pi}\frac{F^{T}}{M_{s}}
 =\frac{1}{{\cal O}(10)}\times \Lambda\,.
 \label{eqn33}
 \end{equation}
%%%%%%%%%%%%%%%%%%%%%%%%%%%%%%%%%%%%%%%%%%%%%%%%%%%%%%%%%%%%%%%%%%%%%

Thus, the contributions to the
$B\mu$ term arises from the loop diagram shown in Fig. \ref{fig1} and also
from the seesaw-type mechanism depicted in Fig. \ref{fig2}.
It is important for us to study whether or not the loop contribution 
is dominated by the seesaw-type contribution.
From Eqs. (\ref{eqn30}) and (\ref{eqn32}),
the ratio of the magnitudes of these contributions is given by
%%%%%%%%%%%%%%%%%%%%%%%%%%% Eqn.34 %%%%%%%%%%%%%%%%%%%%%%%%%%%%%%
 \begin{equation}
  \frac{B\mu^{(\rm loop)}}{B\mu^{(\rm seesaw)}}=
  {\cal O}\left(\frac{1}{10}\;{\rm GeV}\right)\times
  \frac{p}{m^{2}}=\frac{p}{p_{0}}\,,
  \label{eqn34}
 \end{equation}
%%%%%%%%%%%%%%%%%%%%%%%%%%%%%%%%%%%%%%%%%%%%%%%%%%%%%%%%%%%%%%%%%
where $p_{0}$ is defined as
%%%%%%%%%%%%%%%%%%%%%%%%%%% Eqn.35 %%%%%%%%%%%%%%%%%%%%%%%%%%%%%%
 \begin{equation}
 p_{0}={\cal O}(10\; {\rm GeV}^{-1})\times m^{2}\,.
 \label{eqn35}
 \end{equation}
%%%%%%%%%%%%%%%%%%%%%%%%%%%%%%%%%%%%%%%%%%%%%%%%%%%%%%%%%%%%%%%%%
To investigate the cases of different dominant contributions 
to the $B\mu$ term, let us consider the following three cases of 
$p/p_{0}$ separately.
\\ \\
${\bf Case\;(i)}$ $p\sim p_{0}$

Here, the two contributions to the $B\mu$ term are comparable.
Putting $\Lambda \sim{\cal O}(10^{4}\;{\rm GeV})$, 
we have $B\mu^{(\rm loop)}\sim B\mu^{(\rm seesaw)}\sim \mu^{2}$ and
$m_{\lambda_{2}}\sim{\cal O}(1\;{\rm TeV})$.
Thus the $\mu$-problem can be solved. 
This case is phenomenologically viable.
\\ \\
${\bf Case\;(ii)}$ $p\gg p_{0}$

Here, the contribution of loop diagram is dominant. 
Putting $\Lambda \sim{\cal O}(10^{4}\;{\rm GeV})$, 
we have $B\mu^{(\rm loop)}\sim \mu^{2}\gg B\mu^{(\rm seesaw)}$ and
$m_{\lambda_{2}}\sim{\cal O}(1\;{\rm TeV})$.
In this case the $\mu$-problem can be solved.
\\ \\
${\bf Case\;(iii)}$ $p\ll p_{0}$

Here, the contribution of seesaw-type mechanism is dominant.
However, the condition 
$B\mu^{(\rm seesaw)}\sim \mu^{2}\gg B\mu^{(\rm loop)}$ implies
$\Lambda\ll {\cal O}(10^{4}\;{\rm GeV})$.
Then, from Eq. (\ref{eqn33}), we obtain 
$m_{\lambda_{2}}\ll {\cal O}(1\;{\rm TeV})$.
This case is phenomenologically unacceptable.
\\ 

From the above consideration, we find that 
the contribution to the $B\mu$ term depends on 
$p/p_{0}$ (or $p/m^{2}$) and that the contribution of the seesaw-type
mechanism cannot become dominant.
The magnitude of $m$ is controlled by the exponent $b$,
which is closely related to the flavor symmetry.
We now discuss the issue of phenomenologically acceptable values of $b$.
From Eqs. (\ref{eqn14}), (\ref{eqn20}) and (\ref{eqn35}) we have
%%%%%%%%%%%%%%%%%%%%%%%%%%%%%%%%%% Eqn.36 %%%%%%%%%%%%%%%%%%%%%%%%%%%%
 \begin{equation}
  m={\cal O}(10^{18-2b}\;{\rm GeV})
  \label{eqn36}
 \end{equation}
%%%%%%%%%%%%%%%%%%%%%%%%%%%%%%%%%%%%%%%%%%%%%%%%%%%%%%%%%%%%%%%%%%%%%%
and
%%%%%%%%%%%%%%%%%%%%%%%%%%%%%%%%%% Eqn.37 %%%%%%%%%%%%%%%%%%%%%%%%%%%%
 \begin{equation}
  \frac{p}{p_{0}}={\cal O}(10^{4b-20})\,,
  \label{eqn37}
 \end{equation}
%%%%%%%%%%%%%%%%%%%%%%%%%%%%%%%%%%%%%%%%%%%%%%%%%%%%%%%%%%%%%%%%%%%%%%
where we take 
$\langle\overline{S}\rangle=\langle S_{0}\rangle\sim
{\cal O}(10^{17}\;{\rm GeV})$
and $M_{s}\sim{\cal O}(10^{18}{\rm GeV})$.
Therefore, we can obtain phenomenologically viable 
values of $m$ and $p_{0}$ for non-negative integers $b$.
For $b\geq 8$, the assumption $|m|\gg |\mu|$ is not satisfied.
For $b=1- 4$, we have $p_{0}\gg p$, which corresponds to
${\bf case\;(iii)}$.
In the case that $b=5,6$ or $7$, we obtain phenomenologically acceptable
solutions to the $\mu$-problem, which correspond to 
${\bf case\;(i)}$ or $\bf case\;(ii)$.
If $b=5$, then both the seesaw-type mechanism and loop diagram
are dominant contributions to the $B\mu$ term.
If $b=6$ or $7$, the dominant contribution is only the loop diagram.

Another phenomenologically viable solution to 
the $\mu$-problem has been proposed by Giudice and Masiero (G-M).\cite{GM}
In the G-M mechanism, both $\mu$ and $B\mu$ terms are generated
by the K\"ahler potential with non-renormalizable interactions.
Contrastingly, in the present model the K\"ahler potential is assumed to be 
of the minimal form.
Consequently, the G-M mechanism does not act, and, instead,
$\mu$ and $B\mu$ terms come from the superpotential.
In conventional models with the minimal K\"ahler potential 
containing the moduli, both the $\mu_{(\rm gra)}$ term and
the $B\mu_{(\rm gra)}$ term are generated by gravitational 
effects,\cite{moduli} and we have $B\mu_{(\rm gra)}\sim\mu^{2}_{(\rm gra)}$.
From the first term in Eq. (\ref{eqn11}), we then find 
$\mu_{(\rm gra)}\sim \partial_{U}h\;F^{U}x^{s}\langle S_{0}\rangle$,
where $U$ is a complex structure moduli and
$x\equiv (\langle S_{0}\rangle\langle\overline{S}\rangle)/M^{2}_{s}$.
By contrast, in the present model, since we make the
assumption $F^{U}=0$,
neither the $\mu$ term nor the $B\mu$ term can be generated by
gravitational interactions.
Thus, the $B\mu$ term comes only from the loop diagram and 
the seesaw-type mechanism.

In the present model, 
the seesaw-type mechanism also generates the soft SUSY-breaking Higgs 
masses $m^{2}_{H_{u0}}$ and $m^{2}_{H_{d0}}$,
which contribute to the diagonal element $x$ of Eq. (\ref{eqn26}).
Under the condition Eq. (\ref{eqn21}),
the magnitude of $x$ can be approximated as
%%%%%%%%%%%%%%%%%%%%%%%%%% Eqn.38 %%%%%%%%%%%%%%%%%%%%%%%%%%%
\begin{equation}
x\sim \mu^{2}-2\frac{m^{2}}{p}\mu
=\mu^{2}\left(1-\frac{1}{{\cal O}(10^{3})}\times\frac{p_{0}}{p}\right)\,,
\label{eqn38}
\end{equation}
%%%%%%%%%%%%%%%%%%%%%%%%%%%%%%%%%%%%%%%%%%%%%%%%%%%%%%%%%%%%%%
where we have used Eq. (\ref{eqn35}).
The first term here comes from the supersymmetric $\mu$ term, and 
the second term from the seesaw-type mechanism.
{\bf Cases (i)} and {\bf (ii)} are of interest here.
Note that the second term is significantly smaller than the first term.
Thus, the contribution of seesaw-type mechanism is negligible. 
On the other hand, since anti-generation fields behave as messengers,
as in the gauge mediation scenario,\cite{GMSB}
two-loop diagrams through gauge interactions generate
soft scalar masses $m_{0}$ of the same order as the gaugino masses: 
\cite{ourmodel}
%%%%%%%%%%%%%%%%%%%%%%%%%%%% Eqn.39 %%%%%%%%%%%%%%%%%%%%%%%%%%%%
\begin{equation}
 m_{0}\sim m_{\lambda}\sim \frac{g^{2}}{8\pi}\frac{F^{T}}{M_{s}}
 =\frac{1}{{\cal O}(10)}\times \Lambda\,.
 \label{eqn39}
\end{equation}
%%%%%%%%%%%%%%%%%%%%%%%%%%%%%%%%%%%%%%%%%%%%%%%%%%%%%%%%%%%%%%%%
Therefore, in the {\bf cases (i)} and {\bf (ii)} with 
$\Lambda\sim {\cal O}(10^{4}\;{\rm GeV})$, 
the contributions to the soft scalar masses of the two-loop diagrams 
through gauge interactions become dominant.

\section{Conclusion}

In conclusion, we have proposed a phenomenologically viable solution to the
$\mu$-problem in the framework of SUSY-breaking mediation through
anti-generation fields.
In supergravity, a solution to the $\mu$-problem can be 
derived by approximating the scalar potential by using a special form 
of the K\"ahler potential.
We explored an alternative approach to the $\mu$-problem 
that does not involve gravitational interactions.
We found that the $B\mu$ term is generated by a seesaw-type mechanism 
as well as by a loop diagram through 
gauge interactions and that the solution to the $\mu$-problem is closely
related to the flavor symmetry of the model.
In the present model, 
although the contribution of the seesaw-type mechanism cannot become
dominant, the relative magnitudes of the two contributions to the
$B\mu$ term depend on the charge assignment of the flavor symmetry
to the fields.
To find a solution to the $\mu$-problem 
in the framework of string-inspired model,
we need to consider contributions of the superpotential, 
in addition to the K\"ahler potential.
Moreover, the discrete symmetry (flavor symmetry) as a stringy symmetry 
places strong constraints on the low energy physics, such as 
the fermion mass hierarchies, mixing angles, and so on.
We would like to emphasize that the discrete symmetry
plays an important role also in solving the $\mu$-problem.
Furthermore, the seesaw-type mechanism also generates the soft Higgs masses.
However, we find that this contribution is negligible.

The model considered here may seem to regard only a particular case.
However, general perturbative string theories contain both matter fields of
the fundamental representation (generation fields) and those of the
anti-fundamental representation (anti-generation) under the gauge group.
The numbers of generation and anti-generation fields correspond to 
the Hodge number in a compact manifold.
Along the path to low energy physics, the generation fields and
anti-generation fields of sets of vector-like multiplets decouple, and the 
remaining generation fields lead to the MSSM at low energy.
From the viewpoint of string phenomenology, the present model is not
specific to a particular case, 
and it sheds light on an important role of anti-generation
fields.
%%%%%%%%%%%%%%%%%%%%%%%  ACKNOWLEDGEMENTS  %%%%%%%%%%%%%%%%%%%%%%%
\section*{Acknowledgments}
I would like to thank S. Kitakado (Nagoya Univ.) and
T. Matsuoka (Koggakan Univ.) for careful reading of manuscript and for useful
discussions and comments. 
%%%%%%%%%%%%%%%%%%%%%%%%%%%%%%%%%%%%%%%%%%%%%%%%%%%%%%%%%%%%%

\end{document}